\documentclass[fleqn,usenatbib]{mnras}

\usepackage{newtxtext}
\usepackage{todonotes}

\usepackage{ragged2e}

\usepackage{multirow}

\usepackage[T1]{fontenc}
\usepackage{soul}
\newcommand{\kms}{{\,\rm km/s}}
\newcommand{\yr}{{\,\rm yr}}
\newcommand{\au}{{\,\rm au}}
\newcommand{\gyr}{{\,\rm Gyr}}
\newcommand{\myr}{{\,\rm Myr}}
\newcommand{\msun}{{\,\rm M_\odot}}

\newcommand{\zsun}{{\,\rm Z_\odot}}

\newcommand{\nbpp}{\texttt{Nbody6++GPU}}
\newcommand{\mob}{\texttt{MOBSE}}

\newcommand{\Ms}{M$_\odot $}

\newcommand{\rev}[1]{{#1}}

\DeclareRobustCommand{\VAN}[3]{#2}
\let\VANthebibliography\thebibliography
\def\thebibliography{\DeclareRobustCommand{\VAN}[3]{##3}\VANthebibliography}

\usepackage{graphicx}	
\usepackage{amsmath}	
\usepackage{amssymb}	

\title[GW mergers in low-mass cluster triples]{Compact Object Mergers in Hierarchical Triples from Low-Mass Young Star Clusters}

\author[A. A. Trani et al.]{
Alessandro A. Trani$^{1,2}$\thanks{E-mail: aatrani@gmail.com},
Sara Rastello$^{3,4}$,
Ugo N. Di Carlo,$^{3,4,5}$,
Filippo Santoliquido$^{3,4,5}$,
Ataru Tanikawa$^{1}$,
\newauthor{Michela Mapelli$^{3,4,5}$}
\\ 
$^{1}$Department of Earth Science and Astronomy, College of Arts and Sciences, The University of Tokyo, 3-8-1 Komaba, Meguro-ku, Tokyo 153-8902, Japan\\
$^{2}$Okinawa Institute of Science and Technology, 1919-1 Tancha, Onna-son, Okinawa 904-0495, Japan\\
$^{3}$Dipartimento di Fisica e Astronomia `G. Galilei', University of Padova, Vicolo dell'Osservatorio 3, I--35122, Padova, Italy\\
$^{4}$INFN, Sezione di Padova, Via Marzolo 8, I--35131, Padova, Italy \\
$^{5}$INAF-Osservatorio Astronomico di Padova, Vicolo dell'Osservatorio 5, I--35122, Padova, Italy \\}

\date{Accepted XXX. Received YYY; in original form ZZZ}

\pubyear{2021}

\begin{document}
\label{firstpage}
\pagerange{\pageref{firstpage}--\pageref{lastpage}}
\maketitle

\begin{abstract}
A binary star orbited by an outer companion constitutes a hierarchical triple system. The outer body may excite the eccentricity of the inner binary through the von~Zeipel-Lidov-Kozai (ZLK) mechanism, triggering the gravitational wave (GW) coalescence of the inner binary when its members are compact objects.
Here, we study a sample of hierarchical triples with an inner black hole (BH) -- BH binary, BH -- neutron star (NS) binary, and BH -- white dwarf (WD) binary, formed via dynamical interactions in low-mass young star clusters. Our sample of triples was obtained self-consistently from direct $N$-body simulations of star clusters which included up-to-date stellar evolution.
We find that the inner binaries in our triples cannot merge via GW radiation alone, and the ZLK mechanism is essential to trigger their coalescence. Contrary to binaries assembled dynamically in young star clusters, binary BHs merging in triples have preferentially low mass ratios ($q \simeq 0.3$) and higher primary masses ($m_{\rm p} \gtrsim 40 \msun$). 
We derive a local merger rate density of $0.60$, $0.11$ and $0.5 \yr^{-1} \,\rm Gpc^{-3}$ for BH-BH, BH-NS and BH-WD binaries, respectively.
Additionally, we find that merging binaries have high eccentricities across the GW spectrum, including the LIGO-Virgo-KAGRA (LVK), LISA, and DECIGO frequencies. About 7\% of BH-BH and 60\% of BH-NS binaries will have detectable eccentricities in the LVK band. Our results indicate that the eccentricity and the mass spectrum of merging binaries are the strongest features for the identification of GW mergers from triples.

\end{abstract}




\section{Introduction}

In the past six years, more than $50$ binary \rev{compact object (CO)} mergers were detected during the first three observing runs of Advanced LIGO and Virgo gravitational-wave (GW) interferometers \citep{VIRGOdetector,LIGOdetector,abbottGW150914,abbottastrophysics,abbottO1,abbottGW170817,abbottO2,abbottO2popandrate,abbottO3a,abbottO3popandrate,abbottGW190425,abbottGW190412,abbottGW190814,LVK2021BHNS,gwtc-2.1,gwtc-3}. So far, the sample includes the merger of 47 binary black holes (BBHs), 2 double neutron stars and two black hole -- neutron star (BHNS) binaries. Understanding the formation and the merger of binary COs and searching for distinctive signatures of different formation scenarios is of utmost importance to help us interpret the current and future detections of the LIGO-Virgo-KAGRA collaboration (LVK).

Among the main proposed formation channels for merging binary COs, we find: pairing of primordial black holes (BHs) \citep[e.g.,][]{carr1974,carr2016,bird2016,scelfo2018,deluca2021}, isolated binary evolution via common envelope \citep[e.g.,][]{tutukov1973, bethe1998,portegieszwart1998,belczynski2002, belczynski2008,dominik2013, belczynski2016,eldridge2016,stevenson2017,mapelli2017,mapelli2018,mapelli2019, klencki2018,ablimit2018,kruckow2018,spera2019,neijssel2019,eldridge2019}, via stable mass transfer \citep[e.g.][]{kinugawa2014,kinugawa2020,inayoshi2017,vandenheuvel2017,tanikawa2021a,tanikawa2022}, or via chemically homogeneous mixing \citep[e.g.,][]{marchant2016,demink2016,mandel2016,dubuisson2020}, \rev{ dynamical perturbations in the field \citep{micheaely2019,michaely2020}}, dynamical formation in young star clusters (YSCs, e.g. \citealt{banerjee2010,ziosi2014,mapelli2016,askar2017,banerjee2017, rastello2018,banerjee2018,dicarlo2019,dicarlo2020a,dicarlo2020b,kumamoto2019,kumamoto2020,rastello2020,banerjee2021a,trani2021,rastello2021}), globular clusters (GCs, e.g. \citealt{portegieszwart2000,downing2010,tanikawa2013a,rodriguez2015,rodriguez2016,rodriguez2018,samsing2014,samsing2018,zevin2019,antonini2020}), nuclear star clusters (NSCs, e.g. \citealt{oleary2009,miller2009,antonini2012,prodan2015,antonini2016,petrovich2017,gondan2018,rasskazov2019,arcasedda2018,arcasedda2019,arcasedda2020a,arcasedda2020b}) and AGN discs \citep[e.g.,][]{mckernan2012,mckernan2018,bartos2017,stone2017,yang2019,tagawa2020}.

\rev{CO mergers} from hierarchical triple systems were investigated in the context of field triples \citep{antonini2017,silsbee2017,toonen2018,rodriguez2018,vignagomez2021} and triples from binary-binary interactions in globular clusters \citep{antonini2016b,martinez2020,martinez2021,arcasedda2021}. 
In this work we study the evolution of triple systems formed in low-mass young star clusters, focusing on triples composed of COs, either BHs, white dwarfs (WDs) or neutron stars (NSs). 

Unlike previous studies, here we select triples formed self-consistently from $N$-body simulations, which include up-to-date stellar evolution \citep{mapelli2017} and regularized integration scheme for close encounters \citep{kus65}. Our focus is on triples from low-mass ($300$--$1000\msun$) star clusters, which rapidly dissolve within $100 \myr$. All the triples we consider have survived the dissolution of their parent cluster. Our study can be considered complementary to the ones of \citet{kimpson2016,britt2021}, who estimated the merger rate of in-cluster mergers of triples from open clusters.

In Section~\ref{sec:tripleprop} we describe the population of triples that we obtained from our direct-$N$ body simulations of low-mass young star clusters. Section~\ref{sec:GWmergers} discusses the numerical setup we use to follow the dynamical evolution of the triples. Section~\ref{sec:mergobj} presents the properties of merging BBH, BHNS and black hole -- white dwarfs (BHWD) binaries, including their merger rate density and their the mass distribution. Finally, we discuss and summarize our results in Section~\ref{sec:conc}.

\section{Triples' properties}\label{sec:tripleprop}
We select our hierarchical triples from the simulations presented in \citet{rastello2020} who performed a suite of $N$-body simulations using the direct-summation $N$-body code \texttt{NBODY6++GPU} \citep{wang2015} coupled with the population synthesis code \mob{} \citep{mapelli2017,giacobbo2018,giacobbo2018b,giacobbo2018c}.

\subsection{Low-mass star cluster simulations}

\texttt{NBODY6++GPU} is the GPU parallel version of \textsc{nbody6} \citep{aarseth2003} that implements a 4th-order Hermite integrator, Kustaanheimo-Stiefel regularization of close encounters
\citep{stiefel1965,kschain} and individual block time--steps \citep{makino1992}. No post-newtonian terms are included in the version of the code used in \cite{rastello2020}. \mob{} \citep{mapelli2017,giacobbo2018,giacobbo2018b}, is an upgrade  of \texttt{BSE} \citep{hurley2000,hurley2002}, including up-to-date prescriptions for core-collapse supernovae, electron capture, stellar winds, pair instability and pulsational pair instability.
\rev{BH natal kicks are randomly drawn from a Maxwell-Boltzmann distribution with a root mean square velocity of $15 \kms$. The natal kick velocity is reduced by $1-f_{\rm fb}$, where $f_{\rm fb}$ is the fraction of the fallback mass \citep{fry12}. The wind-mass loss rate for massive stars depends on the	electron-scattering Eddington ratio, and considers the increase of the mass-loss rate when a star is close to the Eddington limit \citep{grafner2008,chen2015}. We do not take into account rotationally enhanced mass loss. Pulsational pair-instability supernovae and pair-instability supernovae are treated as in \cite{spera2017} and \cite{mapelli2020}.
All the remaining processes, such as tides, mass transfer, common
envelope and GW orbital decay, are implemented as in \citet{hurley2002}.
The assumptions of our stellar population synthesis model are summarized in Table~\ref{tab:bse}.

The above assumptions crucially affect the evolution of binaries, and therefore the formation and stability of triples in our simulations. In particular, higher-speed natal kicks would likely hinder the formation of stable triples in various ways. First, they would lead to a lowered retention of binaries and COs in the clusters, preventing them to form stable hierarchical triples. Second, natal kicks in stellar triples may lead to dynamical instability, resulting in the disintegration of the triple \citep{pijloo2012,perets2012a,lu2019}.
}

\citet{rastello2020} performed $100002$ direct $N$-body simulations of low-mass young star clusters exploring three different metallicities: $Z = 0.02$, $0.002$ and $0.0002$ ($33334$ simulations per each metallicity). The young star clusters have masses in the range $300\leq{} m_{\rm SC}/{\rm M}_\odot <1000$ sampled from a power-law distribution $dN/dm_{\rm SC}\propto m_{\rm SC}^{-2}$, reminiscent of the distribution of young star clusters in Milky-Way like galaxies \citep{lada2003}. The initial star cluster half mass radius $r_{\rm h}$ is chosen according to \cite{markskroupa12}:
\begin{equation}
r_{\rm h}=0.10^{+0.07}_{-0.04}\,{}{\rm pc}\,{} \left(\frac{m_{\mathrm{SC}}}{{\rm M}_{\odot}}\right)^{0.13\pm 0.04}
\end{equation}

Stellar masses are extracted from a Kroupa \citep{kroupa2001} initial mass function in the mass range $0.1 \le{} m \le{} 150 \msun$.
The orbital parameters of original binaries are set following the distributions of \cite{sana2012}: the binary eccentricities $e$ are randomly drawn from a distribution $p(e)\propto{}e^{-0.42}$ with $0\leq{}e<1$  while the orbital periods $P$ follows the distribution $p(\Pi)\propto{}\Pi^{-0.55}$, where $\Pi\equiv{}\log_{10}(P/\mathrm{days})$ and $0.15\leq{}\Pi\leq{}6.7$. 
The simulations have been performed adopting the rapid core-collapse supernova model \citep{fryer2012}, which prevents the formation of COs in the mass range $2-5$ M$_\odot$. 

The simulated young star clusters initially host 40$\%$ original binaries \footnote{Here and in the following, \emph{original binaries} are stars already bound in a binary in the initial conditions.}.  
Stars are randomly paired by using a distribution $\mathcal{P}(q)\propto{}q^{-0.1}$, where $q=m_2/m_1$ is the ratio of the mass between the secondary and the primary star according to \cite{sana2012}.
Hence, all the stars with mass $\,m\,\ge{}5 \msun$ are  members of binary systems, while stars with mass $m\,<\,5$~\Ms{} are randomly paired until the imposed total binary fraction $f_{\mathrm{bin}}=0.4$ is reached. The result of this method is that the most massive stars (down to $5\msun$) are all binary members, while the fraction of binaries falls to lower values for lighter stars, in agreement with \cite{moe2017}.
The simulated young star clusters are embedded in a solar neighbourhood-like static external tidal field and we put them on a circular orbit around the centre of the Milky Way at a distance $8\,\mathrm{kpc}$ \citep{wang2016}.
Each young star cluster is integrated for a maximum time $t=100\,\mathrm{Myr}$.

\begin{table}
\begin{tabular}{|lc|}
	\hline
	Wind mass loss: $\dot{M}(Z) \propto M^{\beta(Z)}$ & \citet{chen2015} \\
	Supernovae model: rapid core-collapse & \citet{fryer2012} \\
	Pair-instability supernovae & \citet{spera2017} \\
	\multirow{2}*{Common envelope: $\alpha \lambda$ model} & \citet{webbink1984} \\
														   & $\alpha = 5$, $\lambda$: \citet{claeys2014} \\ 
	Natal kicks: $\sigma = 15 \rm\, km/s$ + fallback & \citet{giacobbo2018b} \\
	\hline
\end{tabular}
	\caption{Properties of our binary stellar population synthesis model.}\label{tab:bse}
\end{table}

There are no primordial triples in the initial clusters, meaning that all the hierarchical triples we find in our clusters are dynamically formed  through $4$+body encounters.
We select those triples that have escaped, meaning they have reached a distance from the star cluster's centre larger than twice its tidal radius. 
Moreover, by 100 Myr the clusters' velocity dispersion has lowered down to $0.26$--$0.86 \kms$, and the triples that remain in the clusters have an average velocity of $1.5 \kms$. For this reason, we include in our analysis also the triples that have survived until the end of the integration at 100 Myr. 
Consequently, all our triples have the same age as the clusters, 100 Myr. 
By that time all of the most massive stars have already collapsed into BHs, and most mass loss by stellar winds has already occurred. 
\rev{{We integrated the dynamics of the triples with the direct N-body code \nbpp. Furthermore, we fully took into account the orbital changes induced by stellar and binary evolution of the inner binary and the outer star, because we used our custom version of \nbpp coupled with \mob{} and \texttt{MOSSE} \citep[see][for more details]{dicarlo2019}, respectively. Therefore, important processes relevant to the evolution of triples, such as mass loss and dynamical instability, are treated self-consistently.}
Other processes specific to triples, such as triple common envelope \citep{glanz2021a}, tertiary tides \citep{gao2020} or tertiary mass transfers are not taken into account by \nbpp. However, these processes are important only for close triples \citep{toonen2020}, while the triples in our sample are very wide.}

We refer to \cite{rastello2020} and \cite{rastello2021} for further details on the star cluster simulations.

\subsection{Demography of triples with an inner-CO binary}

In the following, orbital quantities such as semimajor axis $a$ and eccentricity $e$ have the subscripts $_1$ and $_2$ when referred to the inner and outer orbit of the hierarchical triple. Quantities pertaining to the individual bodies, such as the mass $m$, have subscripts $_1$ and $_2$ when referring to the inner binary members, and $_3$ when referring to the outer body.

We first only select triples whose inner binary members are either BHs, NSs or WDs. Table~\ref{tab:triplesum} summarizes the number of triples with an inner BBH, BHWD or BHNS binary for each metallicity set. 
Triples with an inner double neutron star are particularly interesting because of the possibility of characterizing the triple via neutron star pulsations \citep{suzuki2019,suzuki2021}. Unfortunately, we do not find inner double neutron stars in our sample. 
The percentage of triples with an inner CO binary are $0.86\%$, $0.48\%$ and $0.15\%$ at $Z = 0.01, 0.1$ and 1 $\zsun$, respectively. \rev{For brevity, hereafter we call ``inner-CO triples'' all the triples with an inner CO binary, and ``CO triples'' all the triples exclusively composed of COs.}

\begin{figure}
	\includegraphics[width=\columnwidth]{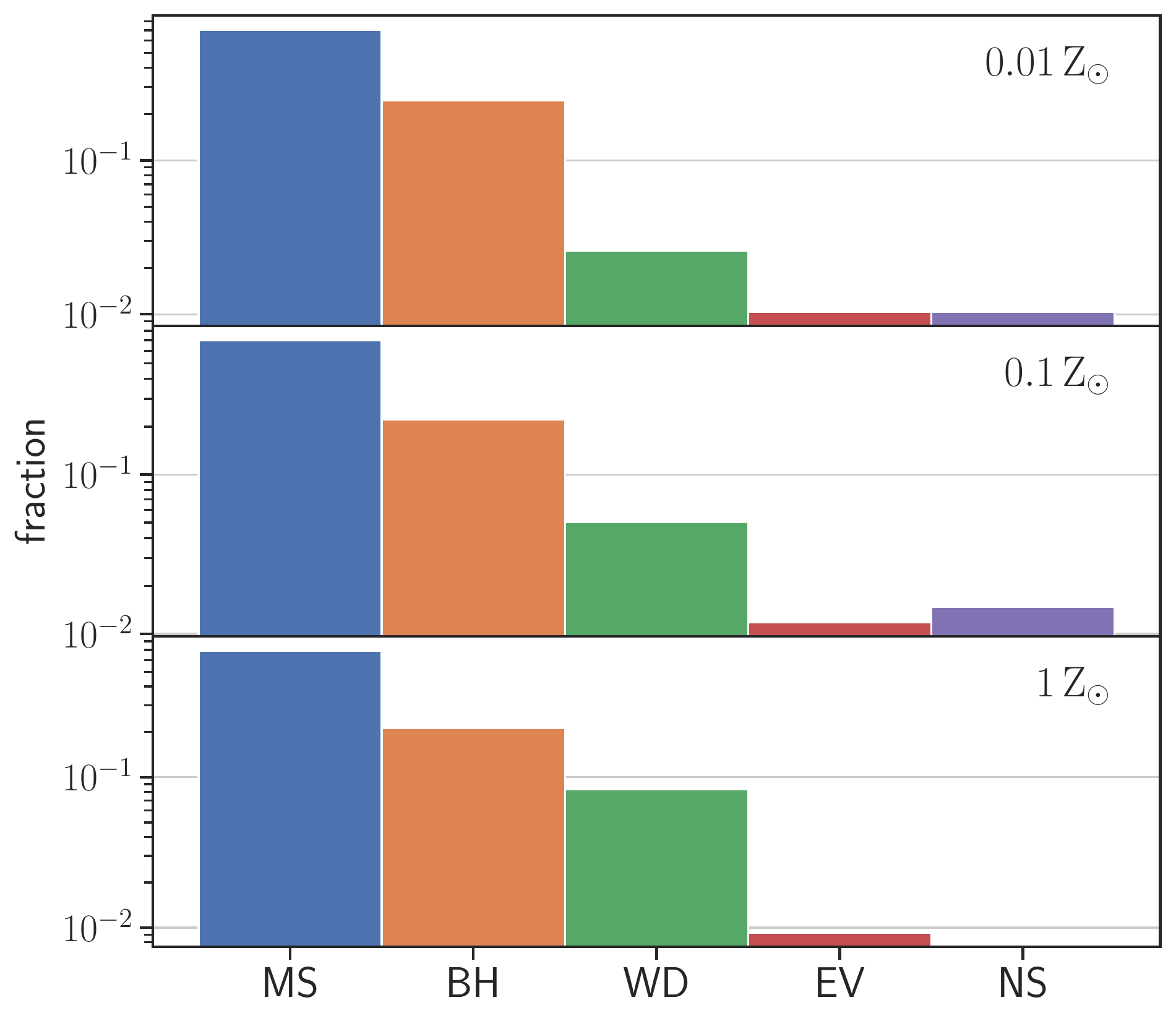}
	\caption{Frequency of the stellar types of the outer object in triples with an \rev{inner-CO} binary. Panels from top to bottom: sets with $Z = 0.01, 0.1$ and  $1 \zsun$. MS: main sequence stars. BH: black holes. WD: white dwarfs. EV: evolved giant stars. NS: neutron stars. Refer to the main text for the precise \textsc{BSE} type each label corresponds to.}
	\label{fig:triplek3}
\end{figure}

In about $25\%$ of the \rev{inner-CO triples}, the outer object is a NS or a BH. The frequency of the remaining stellar types are shown in Figure~\ref{fig:triplek3}, divided per metallicity set. Here, we label as main sequence (MS) objects with \textsc{BSE} type 0 or 1, white dwarf (WD) for \textsc{BSE} types 10, 11 and 12, and evolved star (EV) for \textsc{BSE} types 2, 3, 4, 5 and 6. 

Most of the \rev{inner-CO triples} in our sample have a MS outer companion. In the Universe, such triples may be detected via astrometry measurements of the outer companion \citep{mashian2017,breivik2018,yamaguchi2018,yalinewich2018,shao2019,shikauchi2020,wiktorowicz2020} and subsequently confused as MS--BH binaries. Even so, radial-velocity monitoring might break the observational degeneracy between these two classes of objects \citep{hayashi2020a,hayashi2020b}. However, because the period of the outer binaries is  $1.5 \times 10^4 \yr$ on average, triples formed in low-mass star clusters are too wide to be detected through astrometry. The same consideration applies to \rev{inner-CO triples} with an outer pulsar, which could be detected via pulsar arrival time analysis \citep{hayashi2021}.
\rev{On the other hand, depending on the local environment, wide triples may experience perturbations from flybys and the galactic potential. These may destabilize triples and trigger GW mergers \citep{michaely2020}. We leave this issue to future investigations, and focus on the evolution of the triples in isolation.
}

\begin{figure}
	\includegraphics[width=\columnwidth]{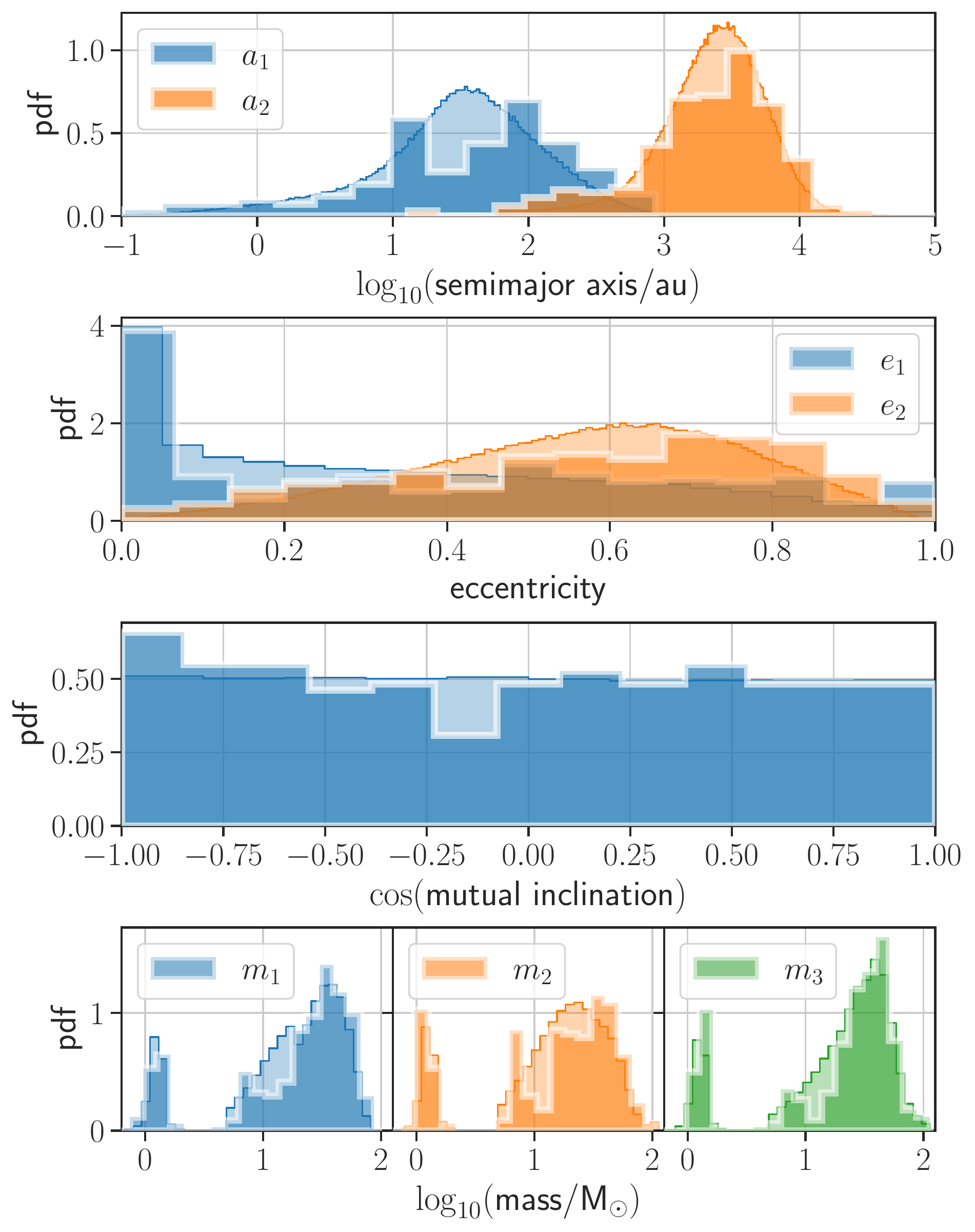}
	\caption{Distributions of initial orbital parameters for the \rev{CO triples}. The dark-shaded histograms indicate the distribution obtained from the original \texttt{NBODY6++GPU} simulations, while the light-shaded histograms were obtained from the Bayesian Gaussian mixture model described in Section~\ref{sec:numsetup}. From top to bottom: semimajor axis of the inner and outer orbit, $a_1$ and $a_2$; eccentricity of the inner and outer orbit, $e_1$ and $e_2$; mutual inclination $i_\mathrm{mut}$; masses $m_1$, $m_2$ and $m_3$. The distributions include all the three sets of metallicities combined.}
	\label{fig:tripleic}
\end{figure}

Figure~\ref{fig:tripleic} shows the masses and orbital parameters of all the \rev{CO triples} in our sample. The median semimajor axes for the inner and outer binary are $\langle a_1 \rangle \simeq 86 \au$ and $\langle a_2 \rangle \simeq 2700 \au$, with a median semimajor axis ratio of $\langle a_2/a_1 \rangle = 30$. 
\rev{An indication of the dynamical origin of our triples is the eccentricity distribution of the outer orbit (Figure~\ref{fig:tripleic}, second panel). At low eccentricity, the distribution grows as a thermal distribution, which is the typical outcome of dynamical interactions \citep{antognini2016,leigh2016c,geller2019}. The cut-off at high eccentricities is linked to the stability of triples: if the outer orbit is too eccentric, the outer star will pass too close to the inner binary, destabilizing the inner orbit and leading to the disruption of the triple.
}

\rev{
In 98\% of all the triples the inner binary was an original binary at the beginning of the $N$-body simulations, indicating that the outer companion was acquired later. Restricting the sample to CO triples only, we find that in 48.5\% of the systems the inner binary was an original binary, in 49.5\% there is no relation between the triples' members, and in the remaining 2\% the outer object and one inner binary member were originally part of an original binary. This indicates that the CO triples, and triples in general, do not form via ``democratic'' binary-binary encounters, but rather via some different mechanisms. In fact, if triples were formed through ``democratic'' encounters, the two original binary members would have similar probabilities to become the inner binary or to break up and one star becoming the outer object. A possibility is that the outer object may be captured through a mechanism analogue to the capture of wide-orbit planets in dispersing clusters \citep{perets2012b}. Another possibility is that that most of the few-body encounters that are producing the triples are not ``resonant'' \citep{hut83a} or ergodic \citep{mon76a,mon76b}. This is unexpected but not unlikely, because the three-body problem is divided into chaotic motion and regular motion \citep{shevchenko2010}, and the statistical theories based on ergodicity can reproduce the results of numerical experiments only after discarding the latter (i.e. flybys and prompt interactions, see for example \citealt{stone2019,manwadkar2020,manwadkar2021,kol2021,ginat2021}).
	
Lastly, half of the CO triples do not contain both members of an original binary in any configuration. Inspecting some of such triples, we find that more than one member results from the merger of an original binary. This suggests that such triples may be formed during 4-body encounters between dynamically formed binaries, or even through more complex 4-body encounters not involving any original binary. Because here we focus mainly on GW sources from triples, we leave the investigation of triple formation mechanisms to future works.
}

\rev{ 
	The presence of ${\sim}50\%$ of inner original binaries in the CO triples raises potentially interesting implications for the spins of COs and their alignment. Specifically, BH spins at merger might not be entirely uncorrelated with the orbital orientation, which has an impact on the spin effective parameter that can be measured from the GW signal. However, in this paper we abstain from making considerations on spin-orbit misalignment for the following two reasons. First, \mob{} does not follow the orientation of the spins in space. While stellar dynamics does not change the spin orientation, it can change the orientation of the binary orbital plane \citep[see][]{trani2021}. Furthermore, the version of \mob{} we used for these simulations does not model the spin of the CO remnants at birth. 
	Second, the inner-CO binaries in our sample are very wide. Because of this, tidal spin-up is inefficient, so any correlation between the binary spins would be due to `primordial' binary spin, whose evolution we cannot follow for the aforementioned reasons.
}

We expect all the triples obtained from the $N$-body simulations to be dynamically stable. We double checked the dynamical stability of the triples using the \citet{mard01} criterion:
\begin{equation}\label{eq:mardaars}
	\frac{a_2}{a_1} > \frac{2.8}{1 - e_2} \left[(1+\frac{1}{q_2}) \frac{1 + e_2}{\sqrt{1 - e_2}}\right]^{2/5} \left(1 - 0.3 \frac{i_\mathrm{mut}}{\pi} \right)
\end{equation}
where $q_2 = (m_1 + m_2)/m_3$ is the mass ratio of the outer orbit. As expected, all the triples were found to be dynamically stable according to the above equation.

If the \rev{inner-CO} binaries were to merge only via GW radiation, their merger time can be estimated using the following expression from \citet{peters1964}:
\begin{equation}\label{eq:delay}
	t_\mathrm{gw} = \frac{15 c^5}{304 G^3} \frac{a_1^4}{(m_1+m_2)\,m_1\,m_2}\,f(e_1)
\end{equation}
where $f(e)$ is a factor that takes into account the orbital eccentricity, which we evaluate numerically as:
\begin{equation}\label{eq:fe}
	f(e_1) = \frac{(1-e_1^2)^4}{e_1^{\frac{48}{19}}(e_1^2 + \frac{304}{121})^{\frac{3480}{2299}}} \int^{e_1}_0 \frac{x^{\frac{29}{19}} (1 + \frac{121}{304}x^2)^{\frac{1181}{2299}}}{(1-x^2)^{3/2}} dx
\end{equation}
 
The median GW merger time for the inner binaries is ${\approx} 2 \times 10^{10} \gyr$, that is an exceedingly long time with respect to the age of the Universe. Only about $2\times 10^{-3}$ of the triples would merge within 13.3 Gyr.
Fortunately, the outer object can shorten the merger timescale of the inner binaries via secular gravitational interactions, which we consider in the next Section.

\begin{table}
	\centering
	\caption{Summary of the triple systems we obtain from the \texttt{NBODY6++GPU} simulations.}
	\label{tab:triplesum}
	\begin{tabular}{lrrrr} 
		\hline
		$Z$ & $N_\mathrm{tot}$ & $N_\mathrm{BBH}$ & $N_\mathrm{BHNS}$ & $N_\mathrm{BHWD}$  \\
		\hline
		$0.01\zsun$ & 67793 & 396 & 101 & 81\\
		$0.1\zsun$ & 70372 & 247 & 47 & 43\\
		$1\zsun$ & 71661 & 76 & 3 & 29\\
		\hline
	\end{tabular}
\begin{flushleft}
	\justify
	Column~1: metallicity ($Z$); 
	column~2: total number of triples ($N_\mathrm{tot}$);
	column~3: number of triples with an inner BBH ($N_\mathrm{BBH}$);
	column~4: number of triples with an inner BHNS ($N_\mathrm{BHNS}$);	
	column~5: number of triples with an inner BHWD ($N_\mathrm{BHWD}$);	
\end{flushleft}
\end{table}

\section{Evolution of hierarchical triple systems}\label{sec:GWmergers}

A hierarchical triple system evolves via secular exchanges of angular momentum between the inner and outer orbits, which drive cyclic oscillations in eccentricities and mutual inclination. This mechanism was originally studied by \citet{lid62} and \citet{koz62} with applications to Earth-orbiting satellites and asteroids in the solar system, respectively. Recently, \citet{ito2019} pointed out a long forgotten work by \citet{zeipel1910} that investigated this mechanism over 50 years before the works of Lidov and Kozai. Therefore, we refer to the secular exchanges of angular momentum in hierarchical triple systems as the von~Zeipel-Lidov-Kozai (ZLK) mechanism \citep[for a review and a book on the ZLK mechanism, see][]{naoz2016,shev2017}.

The ZLK mechanism is particularly important for our problem, because it can drive the eccentricity of the inner binary to extreme values. Together with GW radiation, the eccentricity increase may trigger the coalescence of the inner binary.

To qualify the role of the ZLK mechanism for our triples, we estimate the ZLK timescale as
\begin{equation}\label{eq:kozai}
		T^\mathrm{quad}_\mathrm{ZLK} = \frac{P^2_2}{P_1} \frac{m_1 + m_2 + m_3}{m_3} (1 - e_2^2)^{3/2}
\end{equation}
where $P_1$ and $P_2$ are the periods of inner and outer orbit \citep[e.g.,][]{antognini2015}.
The median ZLK timescale is $T^\mathrm{quad}\approx10 \myr$, which tells us that ZLK oscillations may play an important role in driving the evolution of our triples.

Equation~\ref{eq:kozai} refers only to the quadrupole-level interactions, which correspond to the second order term in the expansion of the three-body Hamiltonian. The next order in this approximation is the octupole-level interaction; this term can cause the inner binary to flip its orientation from prograde to retrograde. During these orbital flips, the inner binary reaches extremely high eccentricity \citep{naoz2013a}. 

Therefore, octupole-level interactions can be crucial in triggering the GW coalescence of the inner binary. We estimate the timescale of octupole-level ZLK oscillations as in \citet{antognini2015}:
\begin{equation}\label{eq:kozai-lidov-octupole}
	T^\mathrm{oct}_\mathrm{ZLK} = \frac{T^\mathrm{quad}_\mathrm{ZLK}}{\sqrt{\epsilon^\mathrm{oct}}}
\end{equation}
where $\epsilon^\mathrm{oct}$ is the ratio of the octupole-to-quadrupole level interaction terms:
\begin{equation}\label{eq:octupleterm}
	\epsilon^\mathrm{oct} = \frac{m_1 - m_2}{m_1 + m_2} \frac{a_1}{a_2} \frac{e_2}{1-e^2_2}
\end{equation}

The octupole-level oscillations that are associated to inner orbit flips occur on a longer timescale with respect to the quadrupole-level oscillations. For our triples,  $T^\mathrm{oct}_\mathrm{ZLK}\approx 3 \gyr$, which is still less than the age of the Universe.

\subsection{Numerical setup}\label{sec:numsetup}
We study the evolution of all triples composed of BHs, WDs or NSs. In this way, we can safely neglect stellar evolution and focus only on triple dynamics.

The sample of \rev{CO triples} that we obtained from the $N$-body simulations is not sufficiently large to obtain a satisfying statistics on the number of GW mergers. We therefore resample the initial conditions using a Bayesian Gaussian mixture model with a Dirichlet process prior \citep{bishop2006}. We apply the model to 7 parameters of the triples: $m_1$, $m_2$, $m_3$, $a_1$, $e_1$, $a_2$, $e_2$. \rev{The mixture model is multivariate and allows us to keep the correlations among these 7 parameters.} \rev{We then sample the arguments of pericenter $\omega_1$, $\omega_2$ and the mutual inclinations $i_\mathrm{mut}$ uniformly in $\cos{(i_\mathrm{mut})}$ between $1$ and $-1$} to obtain the full\footnote{We also sample the longitude of the ascending node $\Omega_1$ in the $[0, 2\pi)$ interval, but its value does not affect the evolution of the triple.} set of parameters needed study the secular evolution of the triples. In this way we are able to preserve the correlations between the properties of the triples.
While resampling the triples, we make sure that they satisfy the stability criterion of Equation~\ref{eq:mardaars}. We also truncate the mass distributions according to the original mass upper limit, to avoid unrealistic mass values that may arise from the tails of the Gaussian mixture.
We generate $10^5$ realizations of \rev{CO triples} per metallicity, for a total of $3\times10^5$ triples. 
Figure~\ref{fig:tripleic} compares the marginal distributions obtained from the Bayesian Gaussian mixture model with the original distributions. The model well reproduces the original distributions, including the low mass gap between BHs and the population of WDs and NSs.

We evolve each triple with the secular evolution code \textsc{okinami}. \textsc{okinami} evolves the double-average, octupole-level equations of motion derived from the 3-body Hamiltonian in Delaunay variables. The equations are integrated with a 7th order Runge-Kutta-Fehlberg integrator with adaptive timestep. \textsc{okinami} includes general relativity precession due to the post-Newtonian term PN1 and GW radiation from the post-Newtonian term PN2.5.

\rev{Adopting the secularly averaged equations allows us to integrate a larger number of triples for a longer time, in contrast with $N$-body methods. However, the secularly averaged equations cannot capture non-secular effects that might be important in modeling CO mergers. Specifically, triples with a weak hierarchy might undergo the so-called quasi-secular evolution, during which the binaries undergo oscillations on a timescale shorter than the secular timescale. This effect is related to the well-known problem of Lunar evection in celestial mechanics, but only recently it was incorporated in the more modern ZLK formalism \citep[see][]{cuk2004,luo2016}. Recent works have shown that quasi-secular evolution may underestimate the binary eccentricities during the GW inspiral \citep{antonini2012,antonini2016}. Consequently, this may overestimate the GW merger times and possibly underestimate the GW merger rates \citep{grishin2018,toonen2018}. The quasi-secular corrections terms for the double-averaged equations have been derived only in a specific reference frame \citep[e.g.][]{luo2016} or in the test-particle approximation \citep{cuk2004}; it is beyond the scope of this paper to derive and implement the correction terms in terms of Delaunay coordinates. For these reasons, our estimates on the merger rates, and especially on the rate of eccentric mergers, should be regarded as a lower limit. 
}

We integrate our triples until either a merger occurs or the total integration time reaches 15 Gyr. A collision happens when $a_1 (1-e_1) < R_1 + R_2$, where $R_1$,$R_2$ are the radii of the inner binary members. For BHs, the radius is set to 50 times the Schwarzschild radius; for NSs we adopt a fixed radius of $10\,\rm km$; for WDs, we use  equation 91 from \citet{hurley2000}. 

It may occur that secular evolution brings the system out of dynamical stability. At each timestep we monitor the stability of the triple using Equation~\ref{eq:mardaars}. If the system does not satisfy the stability condition, we stop \textsc{okinami} and continue the integration using the few-body code \textsc{tsunami} \citep[see e.g.,][]{trani19b}. In order to convert from secular Keplerian orbital elements to Cartesian positions and velocities, we randomly sample the mean anomalies of the inner and outer binaries uniformly in $[0, 2\pi)$. 
We stop the $N$-body integration when the chaotic triple breaks up into an unbound binary-single, or when a merger occurs. In the case of triple breakup, we sum the triple breakup time to the GW merger timescale of the escaping binary, and take it as the merger time.

\section{Properties of mergers}\label{sec:mergobj}

\subsection{Merger times and rates}\label{sec:mergrate}

\begin{figure}
	\includegraphics[width=\columnwidth]{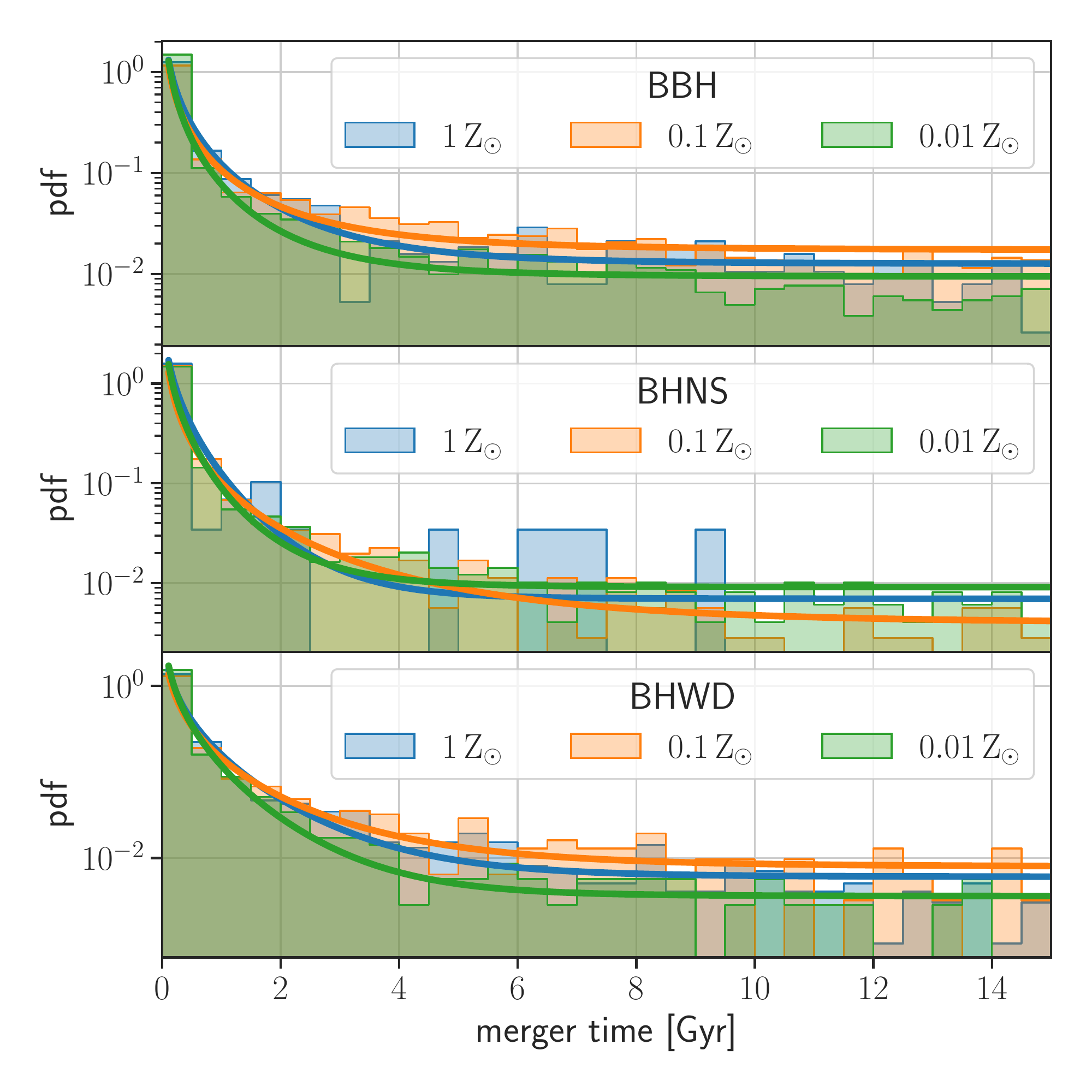}
	\caption{Probability density distribution of the merger time for triples with an inner BBH (top panel), inner BHNS (middle panel) and inner BHWD (bottom panel). The histograms indicate the distribution obtained from the simulations run with \textsc{okinami} and \textsc{tsunami}. The solid lines are the Monte Carlo Markov Chain fit to Equation~\ref{eq:mergertimedist}. Triples merging with the ZLK mechanism have a very short ($<1\gyr$) delay time.}
	\label{fig:mergdist}
\end{figure}

Figure~\ref{fig:mergdist} shows the merger time (also called delay time) distribution for the triples with an inner BBH. The shape of the distribution is very similar for all triples at all metallicities, and  is strongly peaked at ${<} 1 \gyr$. From the merger time distribution, we can expect that triples born at higher redshifts will not contribute much to the local merger rate.
We find that the distributions of delay times are well represented by a mixture model composed of a Weibull distribution plus a flat component:
\begin{equation}\label{eq:mergertimedist}
	p(x) = w_1 \,\frac{\alpha}{\beta} x^{\alpha-1} \exp{\left(-\left(\frac{x}{\beta}\right)^\alpha\right)} + w_0
\end{equation}
The curves in Figure~\ref{fig:mergdist} show the fit to Equation~\ref{eq:mergertimedist} performed with PyMC3, assuming a flat prior on ($\alpha$,$\beta$) and a Dirichlet distribution prior with unitary concentrations on $(w_1, w_2)$. 

We calculate the merger rate density as a function of redshift using the same approach as \citet{santoliquido2020}, in order to compare our results with those of \cite{rastello2020} and \cite{rastello2021}. In particular, the merger rate density in the comoving frame is
 \begin{eqnarray}\label{eq:cosmorate}
   \Gamma_i(z) = f_{\rm YSC} \frac{\rm d\quad{}\quad{}}{{\rm d}t(z)}\int_{z_{\rm max}}^{z}\psi(z')\,{}\frac{{\rm d}t(z')}{{\rm d}z'}\,{}{\rm d}z' \,{}   \nonumber \\
   \int_{Z_{\rm min}(z')}^{Z_{\rm max}(z')}\eta{}(Z)\,{}\mathcal{F}(z',z, Z)\,{}{\rm d}Z,
\end{eqnarray}
where the index $i$ indicates BBHs, BHNSs or BHWDs, $\psi(z')$ is the star formation rate density  at redshift $z'$, $f_{\rm YSC}$ is the fraction of star formation rate that happens in low-mass young star clusters, $t(z)$ is the look-back time at redshift $z$,  $Z_{\rm min}(z')$ and $Z_{\rm max}(z')$ are the minimum and maximum metallicity of stars formed at redshift $z'$, $\eta{}_i(Z)$ is the merger efficiency at metallicity $Z$, and $\mathcal{F}_i(z', z, Z)$ is the fraction of BBHs, BHNSs or BHWDs that form at redshift $z'$ from stars with metallicity $Z$ and merge at redshift $z$, normalized to all BBHs, BHNSs or BHWDs that form from stars with metallicity $Z$. To calculate the look-back time we take the cosmological parameters from \cite{planck2016}.

 The merger efficiency is the total number of BBHs, BHNSs or BHWDs with delay time shorter than the Hubble time, divided by the total initial mass of their host star clusters. For the cosmic star formation rate density, we use the fit from \cite{madau2017}:
 \begin{equation}\label{eq:madau}
\psi{}(z)=0.01\,{}\frac{(1+z)^{2.6}}{1+[(1+z)/3.2]^{6.2}}~\text{M}_\odot\,{}\text{Mpc}^{-3}\,{}\text{yr}^{-1}.
\end{equation}
Finally, 
\begin{equation}\label{eq:Fz}
\mathcal{F}_i(z',z,Z)=\frac{\mathcal{N}_i(z',z,Z)}{\mathcal{N}_{\text{TOT\,{}i}}(Z)}\,{}p(z', Z),
\end{equation}
where $\mathcal{N}_i(z',z,Z)$ is the total number of BBHs, BHNSs or BHWDs that form at redshift $z'$ with metallicity $Z$ and merge at redshift $z$,  $\mathcal{N}_{\text{TOT,\,{}i}}(Z)$ is the total number of BBHs, BHNSs or BHWDs with progenitor's metallicity $Z$, and 
\begin{equation}
\label{eq:pdf}
p(z', Z) = \frac{1}{\sqrt{2 \pi\,{}\sigma_{\rm Z}^2}}\,{} \exp\left\{{-\,{} \frac{\left[\log{(Z(z')/{\rm Z}_\odot)} - {\langle{}\log{Z(z')/Z_\odot}\rangle{}}\right]^2}{2\,{}\sigma_{\rm Z}^2}}\right\},
\end{equation}
is the stellar metallicity distribution at a given redshift. We take the average metallicity $\langle{}\log{Z(z')/Z_\odot}\rangle{}$ from \cite{santoliquido2021} and assume a metallicity spread $\sigma_{\rm Z}=0.2$.

\begin{figure}
	\includegraphics[width=\columnwidth]{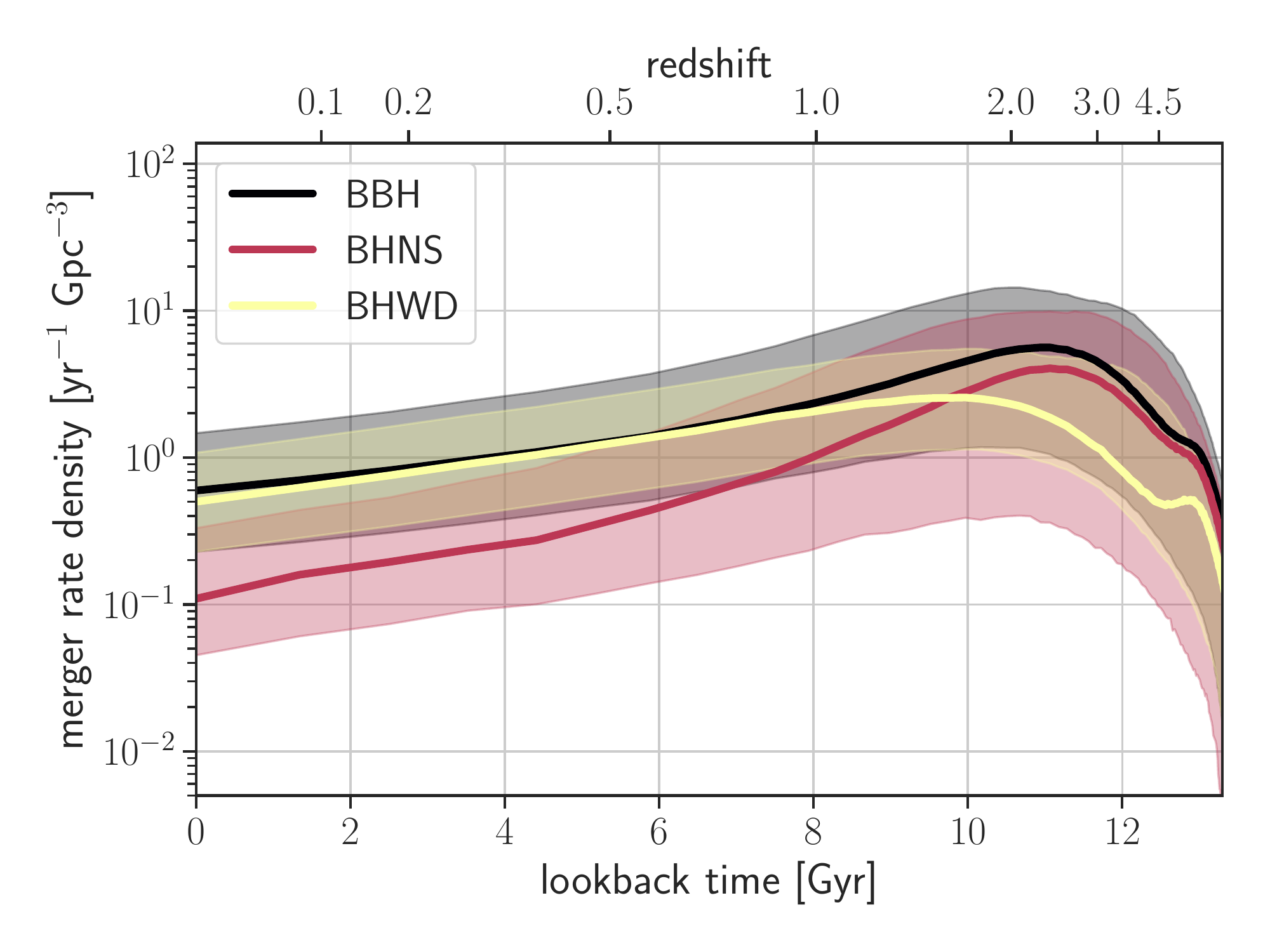}
	\caption{Merger rate density of BBHs (black), BHWDs (yellow) and BHNSs (red) as a function of redshift and lookback time. The shaded area indicates the 90\% confidence interval.
	The merger rate density closely follows the star formation rate density because of the short delay time.}
	\label{fig:rateshift}
\end{figure}

The resulting BBH merger rate density as a function of look-back time (or redshift) is shown in Figure~\ref{fig:rateshift}. As expected from the short delay time, the merger rate closely follows the evolution of the star formation rate density, which peaks at $11 \gyr$ ($z\simeq2$). The largest contribution comes from triples at $0.1 \zsun$, which have only a moderate merger fraction of $f^\mathrm{BBH}_\mathrm{merg} = 0.028$, but can still form at smaller redshifts. The local merger rate density of BBH is therefore:
\begin{equation}\label{eq:bbhmerg}
	\Gamma_\mathrm{BBH} \simeq 0.60^{+0.84}_{-0.37} \,f_{\rm YSC} \,\yr^{-1} \,\rm Gpc^{-3}
\end{equation} 
while for BHNSs and BHWDs, we find 
\begin{equation}\label{eq:bhnsmerg}
	\Gamma_\mathrm{BHNS} \simeq 0.11^{+0.23}_{-0.06} \,f_{\rm YSC} \, \yr^{-1} \,\rm Gpc^{-3}
\end{equation} 
and 
\begin{equation}\label{eq:bhwdmerg}
	\Gamma_\mathrm{BHWD} \simeq 0.50^{+0.59}_{-0.27} \,f_{\rm YSC} \, \yr^{-1} \,\rm Gpc^{-3}
\end{equation} 
respectively.

These merger rates can directly be compared to the ones from dynamically formed binaries, obtained from the same clusters. The binary channel has an expected local merger density rate of ${\sim} 28 \yr^{-1} \rm\,Gpc^{-3}$ for BHNSs \citep{rastello2020} and $88^{+34}_{-26} \yr^{-1} \rm\,Gpc^{-3}$ for BBHs \citep{rastello2021}, which are about 100 times higher than what we have estimated from triples. 

\rev{
	As stated earlier, the final rates depend on the precise value of $f_{\rm YSC}$, which is the fraction of star formation that occurs in clusters similar to the ones we have considered. This value is uncertain, but given an initial cluster mass function we can provide some rough estimates. Assuming a log-uniform mass distribution between 50 and 1000, $f_{\rm YSC} \approx 0.4$ \citep{lada2003}. On the one hand, the final rates may be obtained substituting $f_{\rm YSC} = 0.4$ in (\ref{eq:bbhmerg}), (\ref{eq:bhnsmerg}), and (\ref{eq:bhwdmerg}). On the other hand, this implicitly assumes that (a) the star formation in higher-mass cluster is negligible, and (b) clusters with $50$--$100 \msun$  do not form CO triples, or their merger efficiency is nil.
}

The rates for triples are lower than those of binaries because of the lower merging efficiency. For example, the merger efficiency of dynamically formed BBHs at $Z=0.01\zsun$ is about $1.4×10^{-5} \msun^{-1}$, compared to $2.5×10^{-7} \msun^{-1}$ for triples. The ratio of merger efficiencies between dynamically formed binaries and triples is about 100:1 at all metallicities, for both BBHs and BHNS mergers.

The merger rate might be underestimated because we only simulated triples where all members are COs. In fact, triples with an outer CO only account for 29\% of all the \rev{inner-CO triples} (Figure~\ref{fig:triplek3}). Assuming that the triples with an outer star have a similar evolution as the ones with an outer CO would increment the BBH merger rate to $\Gamma_\mathrm{BBH} \simeq 2.05 \,f_{\rm YSC} \, \yr^{-1} \,\rm Gpc^{-3}$. However, it is unlikely that triples with an outer MS star would lead to a similar merger fraction, because MS stars are less massive than the inner-CO binary, and also because possible mass loss (by stellar winds or mass transfer) would increase the semimajor axis of the outer orbit. 

Our local merger rate densities are similar to those estimated in previous studies on field triples and triples from globular clusters. Specifically, the BBH local merger rate for field triples has been estimated to be $0.14$--$6$ \citep{silsbee2017}, $0.3$--$1.3$ \citep{antonini2017}, and $2$--$23 \yr^{-1} \,\rm Gpc^{-3}$ \citep{rodriguez2018}, while the local merger rate for triples in globular clusters is  $0.4$--$1$ \citep{antonini2016} and $0.35$ $\yr^{-1} \,\rm Gpc^{-3}$ \citep{martinez2020}. 
The various discrepancies among the above studies are to be attributed to different physical ingredients, most importantly the prescriptions for  BH natal kicks, which can greatly affect the survival of \rev{CO triples}. Furthermore, some studies do not consider ZLK evolution during the progenitor stars' lifetime, which likely induces stellar mergers before the \rev{inner binary members can become COs}.

\subsection{Mass distributions}\label{sec:mergprop}

Even though the local merger rates from the triple channel are about 100 times smaller than those from the binary channel, the two channels might be disentangled using other merger properties, such as the masses and the mass ratios.

\begin{figure}
	\includegraphics[width=\columnwidth]{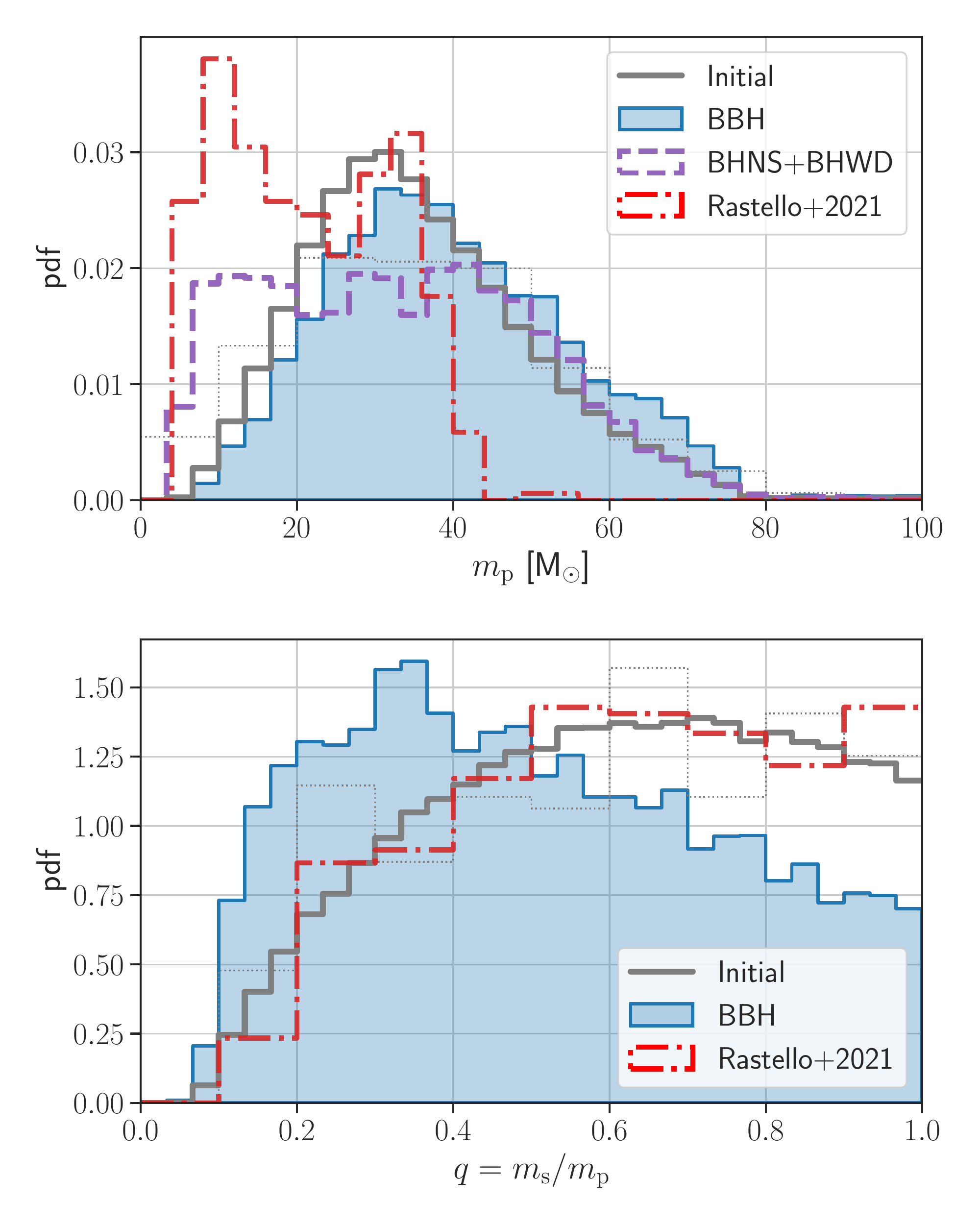}
	\caption{Distributions of primary mass $m_1$(top) and mass ratio $q = m_\mathrm{s} / m_\mathrm{p}$ (bottom) for merging \rev{CO triples}. The different contributions from the three metallicity are weighted according to the local merger rate density. Blue histogram: distributions for merging BBHs. Purple dashed line: distribution for merging BHNSs and BHWDs ($m_1$ distribution only). Grey lines: initial distributions. Red dot-dashed lines: distributions of merging BBH binaries from low-mass young star clusters from \citep{rastello2021}. Thin dotted lines: initial distributions of the original triples from \texttt{NBODY6++GPU}. Each distribution is normalized to unity. The ZLK mechanism in triples favors mergers with lower mass ratios, compared to mergers from binaries.}
	\label{fig:m1_q_dist}
\end{figure}

Figure~\ref{fig:m1_q_dist} shows the distribution of primary mass $m_\mathrm{p}$ and mass ratio $q = m_\mathrm{s} / m_\mathrm{p}$ for merging binaries. 
The distribution of primary masses in merging binaries largely resembles the initial distribution, with small selection effects due to the ZLK mechanism. The primary masses can be substantially higher with respect to binary mergers from the same clusters \citep{rastello2021}. Furthermore, mergers from triples lack the primary mass peak at ${\sim}10 \msun$, which is present in binary mergers and also matches the latest analysis of the GW Transient Catalogue data \citep{gwtc2020b,gwtc-3pop}.

Overall, binaries with massive primaries are more likely to merge than binaries with lower mass primaries, which is not surprising given that the GW coalesce timescale decreases for increasing masses. In addition, binaries with more massive primaries can have smaller mass ratios, which will increase the strength of the octupole ZLK mechanism\footnote{Sometimes referred as the eccentric ZLK mechanism \citep{naoz2016}.}.

This effect can be clearly seen in the bottom panel of Figure~\ref{fig:m1_q_dist}, which shows that mergers occur preferentially at low mass ratios. The importance of the octupole ZLK term for low mass ratio systems was also recently pointed out by \citet{suyubu2021}. 
The comparison between the initial and the merging populations shows clearly that the merger fraction increases at low $q$, with a peak at $q=0.3$. This trend is opposite compared to the distribution of star cluster binaries, which instead follows the initial distribution, which decreases at low $q$.

Our result are in contrast with the results of \citet{martinez2020}, who find no difference in the mass distributions between globular cluster triples and globular cluster binaries. One possible explanation is that triples from globular clusters are more compact, and therefore the inner binaries can merge by GW radiation alone without the need of ZLK mechanism. Therefore, the selection effect on low $q$ and high $m_\mathrm{p}$ is less evident.
Another possible cause of this discrepancy is that our triples originate self-consistently from direct-$N$ body simulations that included stellar evolution and regularized close encounters \citep{wang2015}, while the triples in \citet{martinez2020} are the result of binary-binary scattering in isolation and without stellar evolution effects \citep{fregeau2003}. Additionally, our $N$-body simulations allow us to follow the evolution of in-cluster triples together with the rest of the cluster, while in \citet{martinez2020} the triples do not interact with the rest of the cluster, because of the limitations of the \textsc{cluster monte carlo} code \citep{cmcpaper}.

\subsection{Eccentric mergers}

\begin{figure}
	\includegraphics[width=\columnwidth]{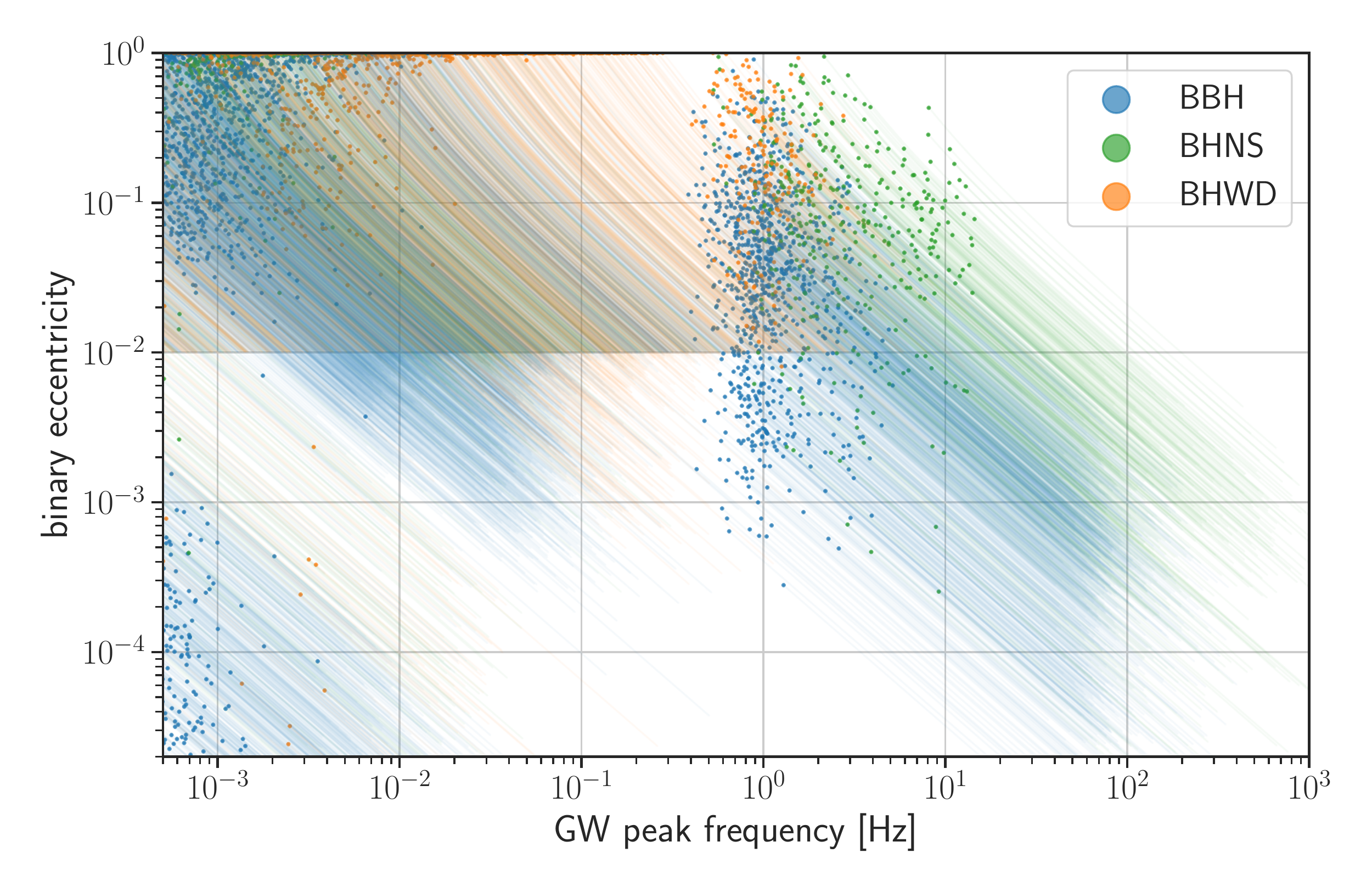}
	\caption{Orbital eccentricity as a function of GW peak frequency for the merging binaries in our simulation. The dots indicate the beginning of the evolutionary track, whose colour depends on whether they are BBHs (blue), BHNSs (green) or BHWDs (orange). As the binaries spiral in, they circularize due to GW radiation.}
	\label{fig:eccfreq}
\end{figure}

Given the extremely high eccentricity during the ZLK oscillations, merging binaries might retain some eccentricity when entering the observable GW bands. While no clear evidence for eccentric mergers exists to date, the LVK network at design sensitivity will be able to distinguish between eccentric and circular inspirals \citep{lower2018,huerta2018,gondan2019}. This will provide a new fundamental piece of information to discriminate among the astrophysical formation scenarios of GWs.
We investigate this by analyzing the last in-spiral phase of the inner binaries in our triple. Figure~\ref{fig:eccfreq} shows the eccentricity of inspiralling binaries as a function of the GW peak frequency, calculated as in \citet{wen2003}. Even though the inspiral phase begins with extremely high eccentricities, most binaries have substantially circularized before entering the LVK band at 10 Hz. At lower frequencies the eccentricity is much higher, especially towards the DECi-hertz Interferometer Gravitational-wave Observatory (DECIGO, peak sensitivity at 0.1 Hz, \citealt{decigo2011,decigo2018}) and the Laser Interferometer Space Antenna (LISA, peak sensitivity at 0.01 Hz) bands, where the eccentricity is close to 1. Note that binaries can enter the LISA band several times before merging, due to repeated ZLK oscillations \citep[e.g.][]{antonini2017,hoang2019,gupta2020}, but here we just display the last part of the inspiral.

\begin{figure}
	\includegraphics[width=\columnwidth]{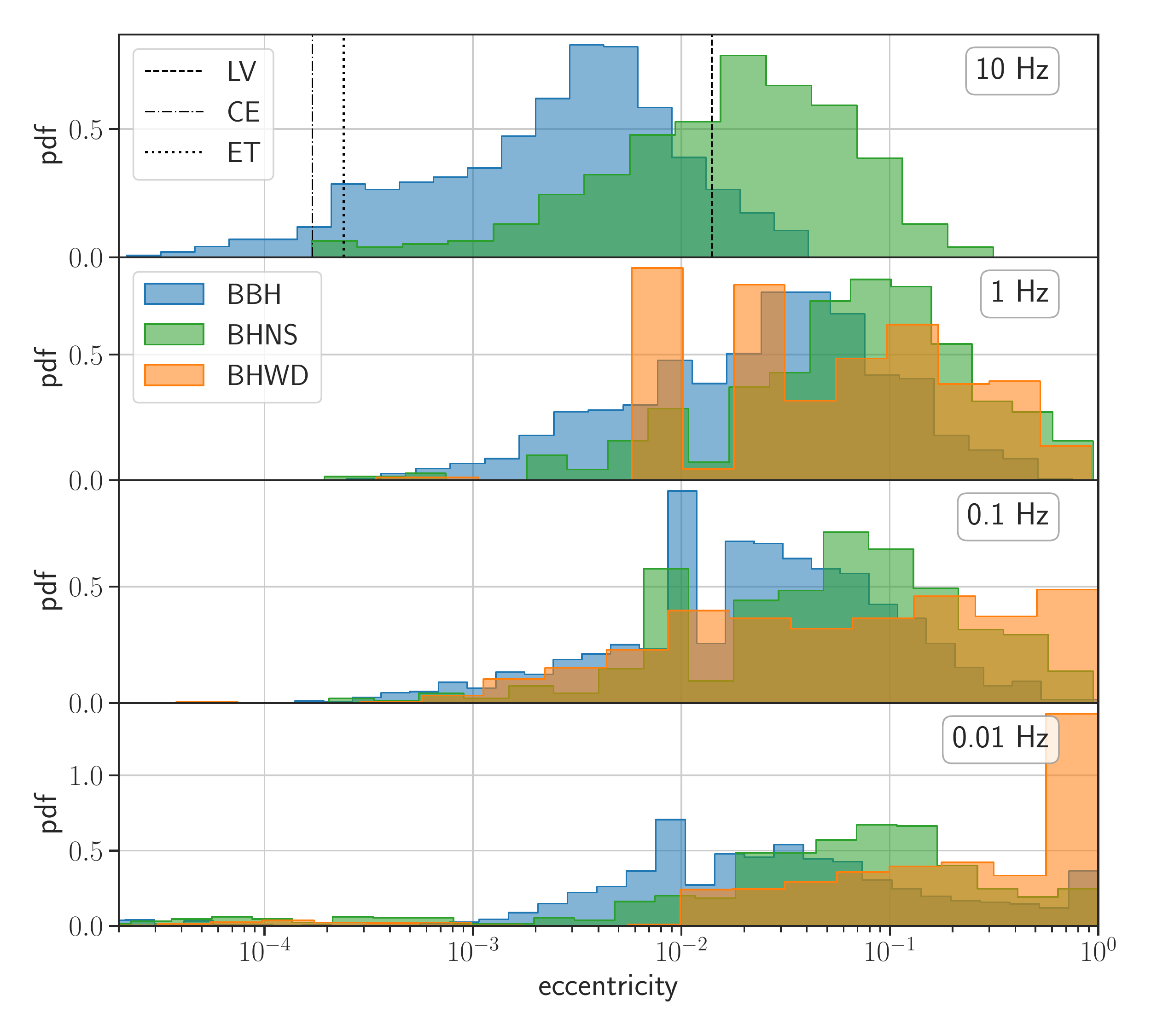}
	\caption{Distributions of orbital eccentricities at different GW peak frequencies. From top to bottom: 10 , 1 , 0.1  and 0.01 Hz. The colours indicate type of  binary: BBH (blue), BHNS (green) or BHWD (orange). Note that BHWD binaries merge before reaching the 10 Hz band. The vertical lines in the top panel indicate the minimum detectable eccentricity for LIGO+Virgo (LV), Einstein Telescope (ET) and Cosmic Explorer (CE) as estimated by \citet{lower2018}.}
	\label{fig:ecchz}
\end{figure}

The distribution of eccentricities at 10 Hz is shown in the top panel of Figure~\ref{fig:ecchz}, divided per binary type. The eccentricity at 10 Hz of BBHs is comparable to that of field triples and triples from globular clusters, whose range is $e_{\rm 10\,Hz} \simeq 10^{-4}$--$10^{-2})$. This also implies that only 7\% of the BBH mergers will have detectable eccentricities in the LVK band. As a caveat, here we may have missed about $10\%$ of highly eccentric inspirals ($1 - e\simeq 10^{-4}$), which emerge when using $N$-body methods, rather than the secular equations \citep{antonini2016b}.

Over 60\% BHNS have eccentricity greater than $e = 0.014$ at 10 Hz. The eccentricity of BHNS binaries is significantly higher, as expected if their merger was driven by the octupole-level interactions. In fact, the BHNS mergers from our cluster triples have similar eccentricities to BBH mergers from in-cluster captures \citep{rodriguez2018b}. This can constitute an important diagnostic to distinguish BHNS mergers from hierarchical triples and those from three-body encounters in young star clusters \citep{rastello2020}. The radius of WDs is much larger than NSs, so that BHWD binaries merge before reaching the 10 Hz band, but can be detected at lower frequencies.

At lower frequencies, the eccentricity distribution shifts to higher values. At ${\sim}1$ Hz, the eccentricities of BBH, BHNS and BHWD is comparable, lying in the range $e_{\rm 1\,Hz}  \simeq 10^{-3}$--$0.5$. As expected, the eccentricity of high mass ratio binaries like BHNS and BHWD is higher than that of BBH. 
This is more manifest at 0.01 Hz, where the eccentricity distribution of all populations becomes bimodal, separated into a low-eccentricity component at $e_{\rm 0.01\,Hz}  \simeq 10^{-3}$--$0.5$ and a high-eccentricity component at $1-e_{\rm 0.01\,Hz}  \simeq 10^{-3}$--$10^{-5}$.

The eccentricity of merging BHWDs and BHNSs determines more than just the GW waveform, because it affects the properties of the merger remnant and of the possible electromagnetic counterpart \citep{fernandez2016,zenati2020}. If the eccentricity at merger is sufficiently high, the merger becomes essentially a head-on collision. For example, head-on collisions between WDs in hierarchical triples have been proposed as a detonation mechanism for type Ia supernovae \citep{raskin2009,hawley2012,katz2012,papish2016}. However, the consequences of low-impact parameter BH--WD and BH--NS collisions have not been explored yet. 

Low eccentricity mergers can instead produce tidal disruption events (TDEs) of WDs. Such WD TDEs might appear as high-energy transients associated with gamma-ray emissions \citep{krolik2011,ioka2016,fragione2020}, and thermonuclear transients \citep{rosswog2009,tanikawa2017,kawana2018,anninos2018}.

\section{Summary and conclusions}\label{sec:conc}

Hierarchical triple systems are composed of a binary orbited by an outer companion star. Such systems are ubiquitous in the Universe, whether they are formed in the field or through dynamical interactions. 
The presence of an outer companion gives rise to secular exchanges of angular momentum between the inner and the outer binary. These exchanges of angular momentum manifest themselves as periodic oscillations in the eccentricity and inclination of the inner binary -- the so called ZLK oscillations.
The increase in the inner binary eccentricity can be the key to explain a variety of observable phenomena. Here, we studied the case in which the increase of the eccentricity  can lead to the rapid coalescence of the inner binary via GW emission.

In this work, we have considered the evolution of hierarchical triple systems with an inner BBH, BHNS or BHWD. The triples were formed self-consistently via dynamical interactions in stellar clusters, which were modeled with direct $N$-body simulations that included up-to-date stellar evolution and accurate integration of close encounters. In particular, we focused on triples from low-mass young star clusters with an initial mass between $300$ and $10^3 \msun$ at three different metallicities ($Z=1$, $0.1$ and $0.01\zsun$, \citealt{rastello2020,rastello2021}). The clusters were evolved up to 100 Myr, at which point they were entirely disrupted by the Galactic tidal field. We then selected the triples composed of COs that were stable and isolation. To obtain a better statistics on CO mergers, we resampled the distributions of the triple's properties using a Bayesian Gaussian mixture model. We finally evolved the triple systems for 15 Gyr using the secularly averaged equations at the octupole-level expansion, including PN1 and PN2.5 corrections for the inner orbit.

We find that ZLK oscillations are a crucial mechanism to trigger the merger of the inner binaries in triples from low-mass young star clusters: only 0.2\% of the binaries would have merged within 13.3 Gyr by GW radiation alone. In contrast, ZLK oscillations cause the inner binaries to merge very quickly: about 90\% of the mergers occur within ${\sim} 1 \gyr$ (Figure~\ref{fig:mergdist}). Because of the short delay times, the merger rate density closely traces the star formation density history (Figure~\ref{fig:rateshift}).

We derive a local merger rate density of $0.60^{+0.84}_{-0.84} \,f_{\rm YSC}\yr^{-1} \,\rm Gpc^{-3}$ for BBHs, $0.11^{+0.23}_{-0.23} \,f_{\rm YSC} \yr^{-1} \,\rm Gpc^{-3}$ for BHNSs and $0.50^{+0.59}_{-0.59} \,f_{\rm YSC} \yr^{-1} \,\rm Gpc^{-3}$ for BHWDs. The rates for BBHs and BHNSs are about 100 times lower than those of binary mergers from the same clusters. \rev{The origin for this difference stems from the merger efficiency of  triple systems, which is about $\sim{100}$ times lower than that of binaries.}

Compared to BBH mergers from open cluster binaries, BBH mergers from triples have more massive primaries, with the $m_\mathrm{p}$ distribution peaked at around ${\sim}30 \msun$ rather than ${\sim}10 \msun$ -- the latter value being more consistent with the latest observational data \citep{gwtc2020b,gwtc-3pop}. 
Another distinctive trait of BBH mergers from triples is the distribution of mass ratios $q=m_\mathrm{s}/m_\mathrm{p}$. In contrast to the cluster binaries pathway, which favours equal mass binaries, the mass ratio distribution for cluster triples peaks at $q = 0.3$ (Figure~\ref{fig:m1_q_dist}). This is caused by the ZLK mechanism, whose eccentricity-pumping effect is enhanced at low mass ratios.

Finally, we show that many BBHs, BHWDs and BHNSs merging through this mechanism will have detectable eccentricities in the LVK, ET and LISA bands. We expect the eccentricity of merging BHNSs in the LVK band to be higher than that of BBHs in the triple scenario (Figure~\ref{fig:ecchz}). The eccentricities in the observable frequencies might be even higher than our estimate,  because we assumed secular approximation, which has been shown to underestimate the eccentricity at merger.

\rev{ Another possible way to identify CO mergers from triple system is from the possible electromagnetic counterpart of BHNS and BHWD mergers. We show that BHWD mergers can occur at both high and low eccentricities, giving rise to TDEs and head-on collisions. The outcome of head-on collisions in BHNS and BHWDs binaries remains to be investigated. }

In conclusion, gravitational interactions in hierarchical triple systems are an important pathway to CO mergers. We have shown that triple systems formed in open clusters can contribute, albeit in a minor part, to the observed GW event rate. Here, we have presented the main properties -- merger rates, masses and eccentricities -- that might help disentangle the origin of present and future events.
In our upcoming work, we will extend our analysis to the \rev{non-CO triples} in our sample, and compare them with field triples.

\section*{Acknowledgements}
We thank Hagai Perets for
a constructive and well thought review.
This work received support from JSPS KAKENHI Grant Numbers 17H06360, 19K03907 and 21K13914.
MM, UNDC, SR and FS acknowledge financial support from the European Research Council for the ERC Consolidator grant DEMOBLACK, under contract no. 770017. AAT would like to thank Naoki Yoshida and Yasushi Suto for insightful discussions. SR thanks Giuliano Iorio for suggestions on data exploration.
\section*{Data Availability}

The \textsc{okinami} code, the initial conditions and the simulation data underlying this article will be shared on reasonable request to the corresponding author.



\bibliographystyle{mnras}
\bibliography{totalms} 








\bsp	
\label{lastpage}
\end{document}